\documentclass[aps, reprint, prb]{revtex4-2}

\usepackage{hyperref}
\usepackage{graphicx}
\usepackage{amsfonts}
\usepackage{gensymb}
\usepackage{amsmath}
\usepackage{physics}
\usepackage{float}
\usepackage{bm}

\usepackage{xcolor}
\hypersetup{
    colorlinks,
    linkcolor={red},
    citecolor={blue},
    urlcolor={blue}
}

\begin{document}

\newcommand{\ingvar}[1]{\textcolor{orange}{#1}}
\newcommand{\jarv}[1]{\textcolor{teal}{#1}}

\title{Hydrodynamic Backflow for Easing the Fermion Sign in Finite-Temperature Electron Path Integral Simulations}

\author{Ingvars Vitenburgs}
\affiliation{Department of Chemistry, Imperial College London, South Kensington Campus, London, SW7 2AZ, UK.}

\author{Jarvist Moore Frost}
\email{jarvist.frost@imperial.ac.uk}
\affiliation{Department of Chemistry, Imperial College London, South Kensington Campus, London, SW7 2AZ, UK.}
\affiliation{Department of Physics, Imperial College London, South Kensington Campus, London, SW7 2AZ, UK.}

\date{\today}

\begin{abstract}

Some notable technology systems, such as high-temperature superconductors and materials for controlled nuclear fusion, require an accurate description of finite-temperature quantum matter. Stochastic path integral methods are finite-temperature and numerically exact, but scale poorly with system size due the notorious Fermion sign problem. To somewhat mitigate this, we use a hydrodynamical backflow coordinate transformation. Our first approach was a continuous normalizing flow machine learning optimisation. We found this to roughly halve the statistical uncertainty at medium sign severity. Numerical issues challenged training effectively. Thus, a semi-analytic analogue was developed to estimate the optimal parameters. We do this by using a derived expression dependent on a Bosonic observable. Hence, the calculation of these values does not have a sign problem. The resulting backflow transformations reduce the problem by multiple orders of magnitude in the specific case of a harmonically trapped, two-dimensional, electron gas at finite-temperature. The total energy of the system agrees with previous, backflow untransformed, studies and we calculate energies for up to 32 electrons. The limiting factor is found to be, primarily, the $O(N^3)$ calculation of the Jacobian, stemming from the coordinate transformation of the backflow. A more thorough implementation may further improve this scaling. Even without this, a route for simulating electron systems at currently unreachable regimes is obtained. 

\end{abstract}

\maketitle

\section{Introduction}

Accurate numerical solution of the Schrödinger equation is at the centrepiece of modern development of materials and molecules. All exact methods fundamentally scale exponentially in the number of electrons. In stochastic methods based upon Monte Carlo integration, this scaling appears as the Fermion sign problem, where taking the arithmetic mean of an increasingly oscillatory integrand leads to an exploding variance.

Stochastic path integral methods are finite-temperature and numerically exact, but scale poorly\cite{Ceperley1991} with system size due to the Fermion sign problem. Some notable technical systems, such as room-temperature superconductors\cite{Pickett2023} and materials for controlled nuclear fusion\cite{Dornheim2024}, require such an accurate description of finite-temperature quantum matter.

In recent decades, stochastic path integral methods for Fermionic matter have been relatively neglected. Recent algorithmic improvements for Bosonic matter\cite{Hirshberg2019} and extrapolation from Bosonic to Fermionic matter via a fictitious non-integer sign\cite{Xiong2022, Dornheim2024} have rekindled interest.
Simultaneously, machine-learning approaches have been developed which offer a more sophisticated trial wavefunction\cite{Thiede2022, Lawrence2024, Xie2022, Schuh2025, Wynen2021}.

Here we apply the most simple (hydrodynamical) backflow coordinate transformation\cite{Feynman1956} to a finite-temperature path integral molecular dynamics simulation of electrons.

\begin{equation}
    \label{equation:Backflow}
    \tilde{\mathbf{r}}(\mathbf{r}) = \mathbf{r} + \sum_{\mathbf{r}_i \neq \mathbf{r}} A \frac{\mathbf{r} - \mathbf{r}_i}{1 + \left( \frac{|\mathbf{r} - \mathbf{r}_i|}{l} \right)^3}.
\end{equation}
We move from the particle position $\mathbf{r}$ to a quasi-position $\tilde{\mathbf{r}}$ based on a collective transformation with a strength $A$, a backflow decay length of $l$ and all of the other particle coordinate $\{ \mathbf{r}_i \}$ values. Though this backflow has a simple closed-form, it has been shown to reduce the errors in zero temperature electron gas simulations\cite{Ros2006}. We came across this form, and the understanding that it could be implemented as a simple modification of independent position to a dependent quasi-position, in the context of its use in describing asymptotic behaviour of the Fermi liquid when modelling non-interacting Fermions\cite{Kruger2008}.

Modifying the open-source path integral molecular dynamics (PIMD) code of Xiong and Xiong\cite{Xiong2022} to include this backflow wavefunction as a direct quasi-position transformation in the path integrals, we then developed methods to optimise the backflow wavefunction parameters.

Chronologically, our first approach was based on a neural ordinary differential equation representation of the transform\cite{Chen2018}, but we found our training implementation fragile and numerically unstable. Our second, analogous, approach directly optimises the backflow parameters using a gradient from a linear expansion (a mismatch term) which we can write down in closed form.
Our approach is somewhat similar to previous work\cite{Holzmann2003, Gantgen2023}, in that we semi-analytically describe the sign error. Our key contribution is that we use a simple approximate form of the \emph{gradient} to optimise the backflow parameters; the approximation does not enter the final energy expression.
We compare this approach to standard gradient-free variational optimisation, and show that for our test systems the results converge to the same basin.

We demonstrate the method on a two-dimensional electron gas in a harmonic trap at finite-temperature with reference data provided by Dornheim\cite{Dornheim2019}. Overall we find that a simple optimised hydrodynamic backflow wavefunction offers a significant improvement in the average sign at finite temperature.

\section{Hydrodynamic backflow in path integral molecular dynamics}

The computational cost of sampling the Fermionic wavefunction has exponential scaling due to the alternating sign in the estimator\cite{Ceperley1991}, a fundamental property of Fermion exchange, $\psi(\dots, r_1, \dots, r_2, \dots) = -\psi(\dots, r_2, \dots, r_1, \dots)$. For Bosonic systems the sign does not change under exchange, and therefore Monte Carlo estimators converge well. Recently, an approach has been developed based on a fictitious partial sign\cite{Xiong2022, Dornheim2024} which extrapolate the behaviour from Boson, to sign-free Boltzmann, and into the Fermion regime. However, these are interpolation based approaches: efforts to directly reduce the Fermionic sign enables one to collect more unbiased data-points from further into the Fermionic region.

Our work was originally motivated by the observation that the sign problem grows as a function of inverse temperature (also particle number; interaction strength), and that the complexity of the nodal surface\cite{Kruger2008} grows as a function of the backflow strength.
Rather than directly try and learn the ground-state wavefunction, we posit that it is more tractable to learn the development of nodal complexity along the progression from non-interacting to fully-interacting system. The \textit{fractal} nature of the wavefunction\cite{Kruger2008, Zaanen2008} suggests that there may be a simple equation of motion which generates the nodal surface, the parameters of which could potentially be inferred from simulations in a relative benign region of parameter space.

We therefore set out to gradually learn a general coordinate transformation $\tilde{\mathbf{r}}(\mathbf{r})$ corresponding to the least noisy average phase $\langle \sigma \rangle$ value.
These type of adjustment coefficients are required for calculating any Fermionic observable by the process of reweighting due to the presence of complex numbers.
More specifically, it can be seen in the expression of an observable average, defined by
\begin{equation}
    \label{equation:Observable}
    \langle O \rangle = \frac{\int \prod_i d\mathbf{r}_i \; O \; | J[\{ \tilde{\mathbf{r}}(\mathbf{r}_i) \}] | \; e^{-S[\{ \tilde{\mathbf{r}}(\mathbf{r}_i) \}]}}{\int \prod_i d\mathbf{r}_i \; | J[\{ \tilde{\mathbf{r}}(\mathbf{r}_i) \}] | \; e^{-S[\{ \tilde{\mathbf{r}}(\mathbf{r}_i) \}]}}.
\end{equation}
Here $S[\{ \tilde{\mathbf{r}}(\mathbf{r}_i) \}]$ is the action, used in the Monte Carlo sampling, $J[\{ \tilde{\mathbf{r}}(\mathbf{r}_i) \}]$ is the Jacobian and the integrand is permuted over the relevant quantum statistics. Hence, it is possible for a $-1 \rightarrow e^{\ln(-1)}$ coefficient to appear in front of the exponent in the case of Fermion statistics, with the $\ln(-1)$ causing the action then to be complex. We can extract this imaginary phase term by using the identity $e^{a + i b} = e^{i b} |e^{a + i b}|$ and infusing it into the observables via $\langle O \rangle \rightarrow \langle O e^{i b} \rangle$. Note that this causes an additional measurement of $\langle e^{ib} \rangle$ from the denominator to be calculated as well. Since the Jacobian modulus $| J[\{ \tilde{\mathbf{r}}(\mathbf{r}_i) \}] |$ in Eq.~\ref{equation:Observable} denominator is positive definite and fairly smooth, we choose to minimise the phase fluctuations only, and so define the average-sign as
\begin{equation}
    \label{equation:Sign}
    \langle \sigma \rangle = \int \prod_i d\mathbf{r}_i \; e^{-S[\{ \tilde{\mathbf{r}}(\mathbf{r}_i) \}]}.
\end{equation}
This phase view of the sign-problem is used in the twist-averaged boundary conditions approach\cite{Lin2001}. If there is no sign problem, the average sign $\langle \sigma \rangle$ approaches unity. Our aim is to maximise the average sign without biasing the simulation, to have the best signal-to-noise in the Monte Carlo estimators.

Previous work\cite{Lawrence2023, Bedaque2017, Alexandru2015} shows that that there indeed exists an optimal maximum phase, found along the best contour for complex-integration.
However, finding this integration contour is challenging, and so far the implementations have been limited to quantum field theory and lattice models.

\subsection{Computational details}

\label{section:ComputationalDetails}

We adapt the open-source implementation\cite{Xiong2022} of path integral molecular dynamics\cite{Hirshberg2020} by Xiong and Xiong to implement the hydrodynamical backflow sampling in Eq. \ref{equation:Backflow}. We also use this code to generate backflow-untransformed learning data $\{ ( \mathbf{r}_i, \sigma_i, T_i ) \}$, in order to find a backflow transformation to maximise the average phase $\langle \sigma \rangle$, as defined in Eq.~\ref{equation:Sign}.
Custom software was written in PyTorch\cite{PyTorch} to process data and optimise the parameters.
The codes used to produce the results reported in this paper are available on GitHub\cite{github} as \textsc{Frost-group/CNF-Sign-Problem}.

We consider the well-studied two dimensional harmonic trap with Coulomb interaction. The Hamiltonian for this system is
\begin{equation}
\label{equation:QuantumDotHamiltonian}
    H = - \frac{1}{2} \sum_i \nabla^2_i + \frac{1}{2} \sum_i \mathbf{r}^2_i + \sum_{i < j} \frac{\gamma}{|\mathbf{r}_i - \mathbf{r}_j|^\alpha}.
\end{equation}
Here the energy scale is the oscillator energy $\hbar \omega$, which effectively describes the width of the trap, and a $\sqrt{\frac{\hbar}{m \omega}}$ length scale. For the Coulomb interaction, $\alpha=1$. Even at medium coupling $\gamma$ and temperature $T$ the sign problem is quite significant\cite{Dornheim2022}.
We use the same settings as reported in the original paper for the codes\cite{Xiong2022}: a total of $5 \cdot 10^6$ steps of $h = 0.0025$ time increment, simulating four particles at $T = 1.0$ temperature, described by the Hamiltonian in Eq.~\ref{equation:QuantumDotHamiltonian} with $\gamma = 0.5$ coupling. At $N = 4$ particles this requires roughly $6$ minutes of single-core compute time, increasing polynomially to $30$ minutes with double the particles and $12$ minutes---linearly---with half the temperature. To cross-check the equivalence of PIMD and path integral Monte Carlo (PIMC) approach, a comparison is given in Tab. \ref{table:XiongDornheimComparison} which displays numerically exact correspondence.

\begin{table}
    \caption{\label{table:XiongDornheimComparison} Comparison of the average sign $\langle \sigma \rangle$ we calculate with the PIMD codes of Xiong and Xiong\cite{Xiong2022} for 2D trapped electrons, described by Eq. \ref{equation:QuantumDotHamiltonian}, compared to those reported from PIMC by Dornheim\cite{Dornheim2019}. 
    The two codes agree within the statistical errors of our sampling.}

    \begin{ruledtabular}
    \begin{tabular}{ccc}
        Parameters & $\langle \sigma \rangle$ & $\langle \sigma \rangle$ Dornheim\cite{Dornheim2019} \\
        \hline
        $N = 3, T = 1.0, \alpha = 1$ & $0.47(1)$ & $0.4746(2)$ \\
        $N = 4, T = 1.0, \alpha = 1$ & $0.24(1)$ & $0.2453(1)$ \\
        $N = 5, T = 1.0, \alpha = 1$ & $0.11(1)$ & $0.1085(1)$ \\
        $N = 6, T = 1.0, \alpha = 1$ & $0.04(1)$ & $0.04184(8)$ \\
        \hline
        $N = 6, T = 0.3, \alpha = 1$ & $0.65(1)$ & $0.6580(2)$ \\
        $N = 6, T = 0.5, \alpha = 1$ & $0.36(1)$ & $0.3616(2)$ \\
        $N = 6, T = 0.8, \alpha = 1$ & $0.11(1)$ & $0.1079(1)$ \\
        $N = 6, T = 1.0, \alpha = 1$ & $0.04(1)$ & $0.04184(8)$ \\
        \hline
        $N = 6, T = 1.0, \alpha = 3$ & $0.51(1)$ & $ 0.5070(1)$ \\
    \end{tabular}
    \end{ruledtabular}
\end{table}

\subsection{Parameters by a neural network optimisation}

\label{section:MachineLearning}

Machine learning techniques have been applied with some success\cite{Hermann2019, Pfau2020} to directly represent the (high-dimensional, ground-state) wavefunction as a neural network, optimised variationally, and thereby significantly improve the accuracy of variational Monte Carlo calculations for small molecules and simple periodic materials. Recently\cite{Thiede2022, Lawrence2024, Xie2022, Schuh2025} the continuous normalizing flow method\cite{Chen2018} has been shown to be able to solve a few simple quantum many-body systems in their ground states as well.

This approach has the advantage of inferring the solution, via fitting a neural-network differential equation, thus being easier to analyse analytically, rather than resulting from a black-box neural network.
In this case, the dynamic of a single point $\mathbf{r}(\sigma, T)$ of some distribution $p[\mathbf{r}(\sigma, T)]$, with respect to the sign value $\sigma$, can be expressed via a representation function set $\{ f_n \}$ with
\begin{equation}
    \label{equation:NormalizingFlowsTransformation}
    \frac{d \mathbf{r}(\sigma, T)}{d\sigma} = \sum_n f_n(\mathbf{r}(\sigma, T)).
\end{equation}

The formalism was designed to describe change in probability distributions, which Lawrence and co-workers have shown makes them apply well to quantum many-body wavefunctions\cite{Lawrence2021}.
In our case, the evolution of the probability distribution can be shown to yield
\begin{equation}
    \label{equation:NormalizingFlowsProbability}
    \frac{d \log(p[\mathbf{r}(\sigma, T)])}{d \sigma} = - \sum_n \; \text{Tr} \left( \frac{d f_n}{d \mathbf{r}} \right).
\end{equation}
This can be used in a \emph{cost} function, after integrating over $\sigma$, to perform maximum likelihood estimation between the model in Eq.~\ref{equation:NormalizingFlowsTransformation} and the data $\{ \mathbf{r}_i(\sigma, T) \}$ with respect to parameters of the $\{ f_n \}$ representation function set.

Previous continuous normalizing flow approaches for solving quantum many-body systems, struggle to choose an appropriate $f_n$ in Eq. ~\ref{equation:NormalizingFlowsTransformation} that does not constrain the number of nodes (spatial manifolds where the wavefunction changes sign).
This is the problem that quantum mechanical wavefunctions can only be considered \emph{quasi}-probabilities.
Hydrodynamic backflow wavefunctions were introduced by Feynman\cite{Feynman1956} for dense Fermion systems, and are now understood as being useful for more general nodal properties of Fermion systems\cite{Kaplis2016}.
They have been demonstrated to house a rich, fractal, nodal structure, with the number of nodes changing continuously as a function of backflow strength.
These backflow forms have also been shown\cite{Ros2006} to stabilise variational and diffusion quantum Monte Carlo studies on electron gases.

To derive the required expressions for this optimisation, we start from the approach of Kruger and Zaanen\cite{Kruger2008}. In a periodic box the non-interacting eigenstates are plane waves, $\phi(\mathbf{k}_n) \sim e^{i \mathbf{k}_n \cdot \mathbf{r}}$. These correspond to a point $\{ \mathbf{r}_i \}$ distribution in coordinate space, after converting via Fourier transform. This is required for application to PIMD methods, which are defined in real space. As an example in one dimension, $r$, and ignoring normalisation factors, this can be shown via
\begin{equation}
\begin{aligned}
    \label{equation:PlaneWaveTransform}
    \phi(r) &= \mathcal{F}_d [e^{i k_n r}]\\ 
    & \propto \sum_{j=1}^\infty e^{i \frac{2 \pi j}{L} r} e^{- i \frac{2 \pi j}{L} r_0} \, \stackrel{L \gg r_0}{=} \, \delta \left( r_0 - r \right).
\end{aligned}
\end{equation}
Here we have removed the periodic constraint by applying $L \gg r_0$, since it is not present, for example, in the quantum dot system above.

These points are then shifted by a collective coordinate transformation via $\{ f_n \}$ from Eq.~\ref{equation:NormalizingFlowsTransformation} and Eq.~\ref{equation:Backflow} yielding
\begin{equation}
\begin{aligned}
    \label{equation:BackflowTransformation}
    \tilde{\mathbf{r}}(\mathbf{r}) &= \mathbf{r} + \sum_n \int_{\sigma_0}^\sigma f_n(\mathbf{r}(\sigma, T)) \; d \sigma
    \\ &= \mathbf{r} + \sum_{n, \mathbf{r}_i \neq \mathbf{r}} \frac{\mathbf{r} - \mathbf{r}_i}{1 + \left( \frac{|\mathbf{r} - \mathbf{r}_i|}{l_n(T)} \right)^3 } \int_{\sigma_0}^\sigma a_n(\sigma, T) \; d \sigma.
\end{aligned}
\end{equation}
Here $\sigma_0$ is the average phase corresponding to distribution data $\{ \mathbf{r}_i(\sigma_0, T) \}$ when no coordinate transformation has been applied---$\tilde{\mathbf{r}}(\mathbf{r}) = \mathbf{r}$ and $\int_{\sigma_0}^\sigma a_n(\sigma, T) \; d \sigma = 0$. It can also be seen that it is more useful to represent the integral $\int_{\sigma_0}^\sigma a_n(\sigma, T) \; d \sigma = A_n(\sigma, T)$ as a neural network rather than the $a_n(\sigma, T)$ function itself.
Hence, the backflow strengths $\{ A_n(\sigma, T) \}$ and length scales $\{ l_n(T) \}$ are now neural networks describing the continuous normalizing flow in Eq. \ref{equation:NormalizingFlowsTransformation}. Their architecture is detailed in Appendix \ref{appendix:NeuralNetworkDetails}, they are learned by back-propagating over individual samples and characterize the nodal surface structure of the Fermionic system.
Furthermore, the summation is over all of the other particle coordinate points $\{ \mathbf{r}_i(\sigma_0, T) \}$ at the same imaginary time step---bead number, in stochastic path integral terminology---of the untransformed distribution. 

While there are\cite{Kwon1993} more sophisticated forms of backflow transformations, their use would make the analysis too complex, whilst perhaps not being more expressive. 
In combination with Eq.~\ref{equation:NormalizingFlowsProbability} we now have a cost function $\sum_i \log(p[\mathbf{r}_i(\sigma, T)])$ for performing maximum likelihood estimation on $\{ \mathbf{r}_i(\sigma, T) \}$ data where
\begin{equation}
    \label{equation:BackflowProbability}
    \log(p([\mathbf{r}(\sigma, T)])) = - \int_{\sigma_0}^\sigma \sum_n \; \text{Tr} \left( \frac{d f_n}{d \mathbf{r}} \right) \; d \sigma .
\end{equation}
Here the prior probability distribution of the simulation data $\log(p([\mathbf{r}(\sigma_0, T)]))$ has been neglected, as it is a normalisation constant and does not affect the optimisation process.

\begin{figure}
    \centering
    \includegraphics[width=1.0\columnwidth]{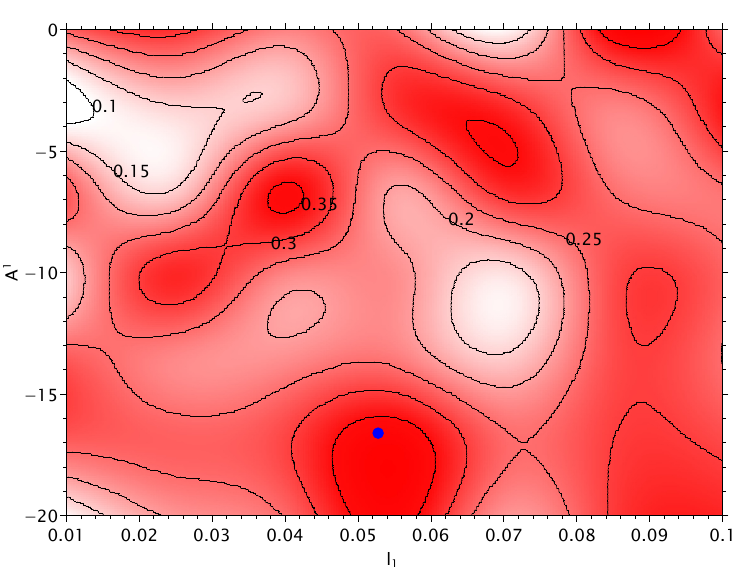}
    \caption{
    Interpolated 8 by 8 uniform grid sampling of the average sign $\langle \sigma \rangle$ with respect to the length $l$ and strength $A$ scales of a single applied backflow in Eq. \ref{equation:Backflow}. This is displayed as a red contour, whilst the blue dot denotes the global maxima of a sign value of $0.5(1)$, obtained by the machine learning approach at $\{A = -15.83, l = 0.05374\}$ described in Appendix \ref{appendix:NeuralNetworkDetails}. In comparison, by this grid sampling, a value of $0.4(1)$ is obtained very close to this point at $\{A = -16.7, l = 0.055\}$. Here the two-dimensional trapped electron Hamiltonian in Eq. \ref{equation:QuantumDotHamiltonian} was simulated with backflow-modified path integral molecular dynamics codes of Xiong and Xiong\cite{Xiong2022}. Temperature $T = 1.0$, electron coupling parameter $\gamma = 0.5$ and $N = 4$ particle parametrisation was used.}
    \label{figure:TestCaseBackflowStrengthAndScale}
\end{figure}

The details of this process is given in Appendix~\ref{appendix:NeuralNetworkDetails} and the main result depicted in Fig.~\ref{figure:TestCaseBackflowStrengthAndScale}---the learned optimal sign backflow strength and length scale values $\{A = -15.83, l = 0.05374\}$ of a single backflow match quite closely the global maximum located by grid-search at  $\{A = -16.7, l = 0.055\}$. 
A sign increase of $\langle \sigma \rangle = 0.2(1) \rightarrow 0.5(1)$ is found as well, with the energy being left unchanged. Hence, we have shown that, in principal, the sign problem can be improved by implementing a hydrodynamical backflow transformation. 

Interestingly, applying multiple backflows (i.e. $n > 1$ in Eq. \ref{equation:BackflowTransformation} summation) did not seem to improve this value further.

Unfortunately, this process, in terms of large-scale application, was found to be unstable in training (see Appendix~\ref{appendix:NeuralNetworkDetails} for further details), and inefficient---batches of converged molecular dynamics coordinate samples were required for learning.
Hence, based on our experience, we therefore turned to developing a semi-analytic optimisation approach. 

\subsection{Semi-analytical optimisation of backflow transforms}

Our aim here was to find an approximate gradient to power standard gradient descent optimisation of the the backflow parameters. 
This is, effectively, a semi-analytic analogue to the machine-learning optimisation undertaken in the previous section.

To explain our derivation, we start with a two-particle Fermion system in one dimension.
The wavefunction, with an applied backflow and centred at the origin, is defined as
\begin{equation}
\begin{gathered}
    \label{equation:TwoBodyBackflowWavefunction}
    \psi(x_1, x_2) =
    \begin{vmatrix}
        e^{i k_1 ( x_1 + \eta[-x_{21}])} &
        e^{i k_1 ( x_2 + \eta[x_{21}])} \\
        e^{i k_2 ( x_1 + \eta[-x_{21}]) } &
        e^{i k_2 ( x_2 + \eta[x_{21}]) }
    \end{vmatrix}, \\
    \eta [x_{21}] = A \frac{x_{21}}{1 + \frac{|x_{21}|^3}{l^3}}, \\
    x_{21} = x_2 - x_1.
\end{gathered}
\end{equation}
Here $A$ and $l$ are the strengths and lengths of a single backflow transformation, as discussed previously. The discretised momentum $k_i = \frac{2 \pi n_i}{L}$ with $n_1 < n_2$ and $n_1, n_2 \in \mathbb{Z}$ represents a periodic box with the unit cell at $x \in [- \frac{L}{2}, \frac{L}{2}]$.
As our work in Sec.~\ref{section:MachineLearning} yielded a very small $l$ length, we take a linear (short-range) approximation with respect to the expansion coefficient, $l^3 \ll 1$.
This results in an adjustment of $1 + \left( \frac{x_1 - x_2}{l} \right)^3 \approx \left( \frac{x_1 - x_2}{l} \right)^3$ to the wavefunction, producing a \emph{linear} correction of the wavefunction, $\psi(x_1, x_2) = \psi_0(x_1, x_2) + \psi_1(x_1, x_2)$, where $\psi_0(x_1, x_2)$ is the unperturbed---null, $l \rightarrow 0$, backflow---case. This, specifically first-order case, corresponds to the form used by Feynman\cite{Feynman1956} to describe correlations in He$^4$ and since been analytically studied\cite{Kruger2008} to produce up to two-particle interactions.

The corresponding energy is from solving the determinant and then integrating the kinetic energy operator $\frac{1}{2} \partial^2_{x_1} + \frac{1}{2} \partial^2_{x_2}$ in coordinate space,
\begin{equation}
\begin{aligned}
    \label{equation:FirstOrderEnergy}
    E_0 + E_1 = & \frac{1}{2} k_1^2 + \frac{1}{2} k_2^2 + \\
    &\underbrace{i \frac{A l^3 (3
    - 2 \gamma_E)}{L^2} (k_1 - k_2)^2 (k_1 + k_2)}_\text{Exchange term}.
\end{aligned}
\end{equation}
Here $\gamma_E$ is the Euler–Mascheroni constant ($\approx 0.5772$) and the convention $k_2 > k_1$ was used to simplify the algebra. 

Note that there is also a \emph{divergent} energy contribution, proportional to $\lim_{\delta \rightarrow 0} [ \log (\delta (k_2 - k_1))$, which has been omitted. 
Our expression was obtained by making the integral, required for deriving Eq. \ref{equation:FirstOrderEnergy}, finite via introducing a $x_2 = x_1 + \delta$ shift of the integration variables. 
Most likely the divergence is caused by the presence of a $x_1 - x_2 \rightarrow 0$ first-order approximation $l^{-3}[l^3 + (x_1 - x_2)^3] \approx l^{-3}[(x_1 - x_2)^3]$ which effectively removes a cut-off $l^3$ for the backflow transformation. 
We expect this to be compensated for by higher order terms, but, due to the complexity of these calculations, we could not verify this even at second order. We really would need the $Al^6$ term in Eq. \ref{equation:FirstOrderEnergy} in closed form.

Because we have disregarded this ultraviolet divergence, our linear expansion is no longer a rigorous physical energy. 
Instead, we only use it as an empirical gradient to optimise the backflow parameters $A$ and $l$, to a region of minimal Fermion sign.

The exchange term flips sign in the Boson case, therefore the \emph{determinant} in Eq.~\ref{equation:TwoBodyBackflowWavefunction} would be replaced by a \emph{permutant}.

The sign can be defined\cite{Xiong2022} as the exponentiated imaginary difference between the Boson $E_b$ and Fermion $E_f$ energies. After a Fourier transform to coordinate space, similar to the procedure in Eq.~\ref{equation:PlaneWaveTransform} and derived in Appendix \ref{appendix:BackflowMomentumSpace}, extension to the many-particle and two-dimensional case, our final, first-order, sign expression is
\begin{equation}
\begin{gathered}
    \label{equation:BackflowSignExpression}
    {\sigma} = e^{i \beta (E_b
    - E_f)} = \cos \Biggl( \frac{A l^3
    \beta (3 - 2 \gamma_E)}{2 \pi^2} \times \\ \times \sum_{i < j} \Biggl[ \frac{2 |\mathbf{r}_j|^3 + 2 |\mathbf{r}_i|^3 + |\mathbf{r}_j - \mathbf{r}_i|^2 |\mathbf{r}_j + \mathbf{r}_i|}{|\mathbf{r}_j|^4 |\mathbf{r}_i|^4} \Bigg] \Biggr) + \\ + i \sin \Biggl( \frac{A l^3
    \beta (3 - 2 \gamma_E)}{2 \pi^2} \times \\ \times \sum_{i < j} \Biggl[ \frac{2 |\mathbf{r}_j|^3 + 2 |\mathbf{r}_i|^3 + |\mathbf{r}_j - \mathbf{r}_i|^2 |\mathbf{r}_j + \mathbf{r}_i|}{|\mathbf{r}_j|^4 |\mathbf{r}_i|^4} \Bigg] \Biggr).
\end{gathered}
\end{equation}
Note that the presence of stand-alone coordinates is due to this essentially being a linear expansion of a much more complex interaction. 
For example, the expansion of $(x_2 - x_1)^{-1} = x_2^{-1} + x_1 x_2^{-2} + O(x_1^2)$ is a trivial example of what we are attempting. 

The appearance of an imaginary energy contribution from a Hermitian operator is surprising, but similar complex shifts appear in other methods to improve sampling\cite{Gantgen2023}, and describing\cite{He2025} phase transitions. 

It is worth noting that the final expression ends up being independent of the hard-confinement length $L$, which is physically sensible for non-periodic and non-confined systems. 
To calculate any higher order---more long-range---energy corrections, perturbation theory can be used by expanding with respect to the expansion coefficient $l^3$. 
More specifically, the effects of the backflow can be interpreted as a perturbation $\delta H$, without needing the expression in closed form.
Even if possible, such an algebraic manipulation would most probably not yield a useful form for physical understanding of the underlying processes.
Therefore we use some ideas from the recursive approach\cite{Taddei2015}.
In this case a low order approximation, such as the one derived in Eq.~\ref{equation:BackflowSignExpression} above, can be applied successively to further optimize the nodal surface sampling, rather than just once (we consider a double application).
Hence, the strength and length minima of the second backflow can be used to find these optimal parameters for the first one.

\begin{figure}
    \centering
    \includegraphics[width=1.0\columnwidth]{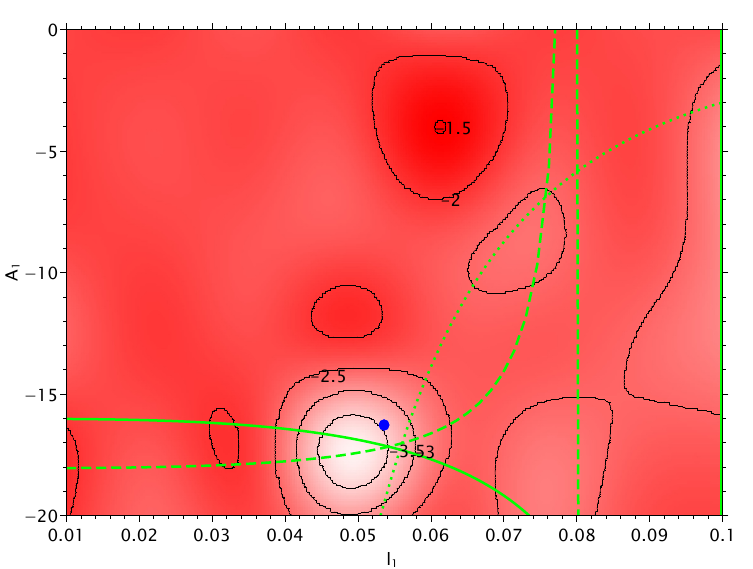}
    \caption{
    Background is the average sign, for given length $l$ and strength $A$ of a single applied backflow. 
    The quantity $\log|\delta A l^3| = |\langle \tilde{A l^3} \rangle - A l^3|$ is plotted as a red contour. 
    The green solid, dashed and dotted curves are plotted using Eq. \ref{equation:BackflowGradient} and Eq. \ref{equation:BackflowCondition} gradients from $\{ A = -10, l = 0.1 \}$, $\{ A = -20, l = 0.08 \}$ and the origin, respectively. The near-intersection of the green lines is close to the blue dot, which is the $\{ A = -15.83, l = 0.05374 \}$ machine learning optimised parameterization described in Appendix \ref{appendix:NeuralNetworkDetails}. This in turn is close to the basin around $\{ A = -17, l = 0.05\}$ of the red contour, which is the global optimum. Here the two-dimensional trapped electron Hamiltonian in Eq. \ref{equation:QuantumDotHamiltonian} was simulated with backflow-modified path integral molecular dynamics codes of Xiong and Xiong\cite{Xiong2022}. Temperature $T = 1.0$ and electron coupling parameter $\gamma = 0.5$ parametrisation was used.}
    \label{figure:TestCaseOptimalBackflowParameters}
\end{figure}

More specifically, the ideal (least noisy) wavefunction samples are taken when $\langle \sigma \rangle$ is at a saddle point of the complex rotation in Eq. \ref{equation:BackflowSignExpression}. 
This means it is either fully real or imaginary with no phase present when measuring observables via Eq. \ref{equation:Observable} and therefore no sign (phase) problem. 
Hence, the condition for $A l^3$ can be shown to be
\begin{equation}
\begin{gathered}
    \label{equation:BackflowCondition}
    \langle A l^3 \rangle = \Biggr\langle - \frac{\pi^3}{\beta (3
    - 2 \gamma_E)} \times \\ \times \Biggr( \sum_{i < j} \frac{2 |\mathbf{r}_j|^3 + 2 |\mathbf{r}_i|^3 + |\mathbf{r}_j - \mathbf{r}_i|^2 |\mathbf{r}_j + \mathbf{r}_i|}{|\mathbf{r}_j|^4 |\mathbf{r}_i|^4} \Biggr)^{-1} \Biggr\rangle.
\end{gathered}
\end{equation}
Note that there is a saddle point at $Al^3 = 0$ as well, but we neglect it here. 
This is because it represents the trivial case when the Fermionic and Bosonic energies are the same, which is not the situation we are interested in. 
Hence, we use the first possible saddle point for Eq. \ref{equation:BackflowSignExpression} at a sign value of
\begin{equation}
    \sigma = \cos \left( \frac{\pi}{2} \right) + i \sin \left(\frac{\pi}{2} \right) = i.
\end{equation}

As an initial test, we calculate the observable in Eq. \ref{equation:BackflowCondition} for the system studied previously.
Indeed, a calculation of this value from a simulation with an applied backflow parameterization $\{ A = -15.83, l = 0.05374 \}$, obtained from the machine learning approach, converged in the same fashion as in Fig.~\ref{figure:TestCaseSignConvergence} and yielded an estimate of $ \langle \tilde{A l^3} \rangle = -2.6 \cdot 10^{-3} \approx A l^3 = -2.4 \cdot 10^{-3}$ with both being quite close to each other.

To confirm that we are converging to the global minimum, direct grid sampling was conducted for the $\log| Al^3 - \langle \tilde{A l^3} \rangle|$ observable in an identical approach to Fig.~\ref{figure:TestCaseBackflowStrengthAndScale} at various strength and scale values and is displayed in Fig.~\ref{figure:TestCaseOptimalBackflowParameters}.
We can see that there is a global minimum, subject to a small binning deviation, as before, further supporting this statement, where the machine learned results have been depicted by a blue dot.
Hence, this provides for a gradient descent, using the first-order perturbation ($l^3 \ll 1$) from Eq.~\ref{equation:BackflowCondition}, where the value $\delta A l^3 = \langle \tilde{A l^3} \rangle - A l^3$ states the optimising direction.

Because the gradient is Bosonic, finding the minimum is sign-problem free.
The calculation of the sign $\langle \sigma \rangle$ itself---which is the reweighting adjustment to Fermi statistics---is not required. 
The addition of a single backflow transformation has a polynomial $O(N^3)$ cost scaling, but this could be\cite{Griewank2000} improved by using more efficient algorithms for Jacobian estimation. 
For our naive implementation, the calculation as at the start of Sec.~\ref{section:ComputationalDetails} with backflow takes $15$ minutes, and $6$ minutes without.

\begin{figure}
    \centering
    \includegraphics[width=0.8\columnwidth]{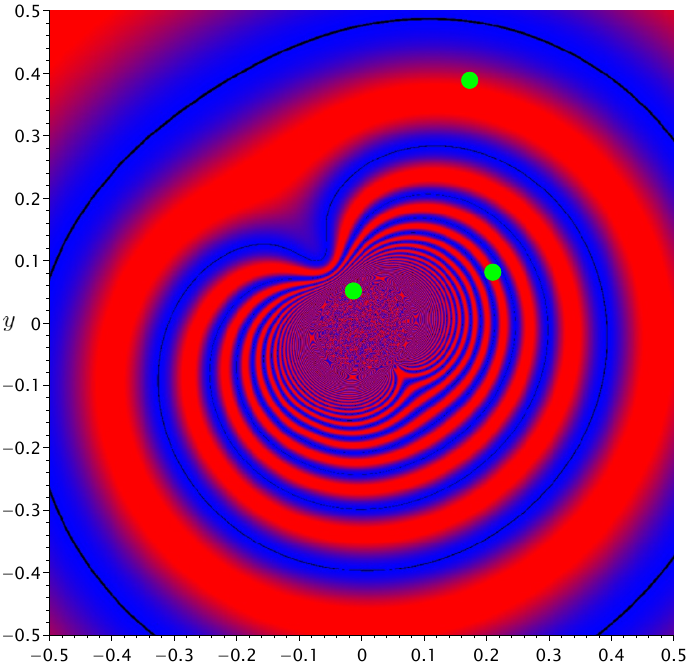}
    \caption{The improved sign sampling distribution of the two dimensional nodal surface, learned from the machine learning process. The green dots represent the position of the other three particles whilst the location of the fourth one is varied to calculate the imaginary part of the sign from Eq. \ref{equation:BackflowSignExpression} with parameters $\{ A = -15.83, l = 0.05374 \}$ described in Appendix \ref{appendix:NeuralNetworkDetails}. The blue denotes a negative sign, whilst the red denotes a positive one. This corresponds to the two-dimensional trapped electron Hamiltonian in Eq. \ref{equation:QuantumDotHamiltonian}. Temperature $T = 1.0$, electron coupling parameter $\gamma = 0.5$ and $N = 4$ particle parametrisation was used.}
    \label{figure:TestCaseSignSampling}
\end{figure}

As an example of this gradient descent method, in Fig.~\ref{figure:TestCaseOptimalBackflowParameters} the magnitudes of the steps $\delta A l^3$ are evaluated at three arbitrarily chosen $A$ and $l$ values $\{ A=-20, l=0.08 \}$, $\{ A = -10, l = 0.1 \}$ and the origin. 
The suggested shifts, from calculating the Bosonic observable in Eq. \ref{equation:BackflowCondition}, were $\delta A l^3 = - 1.0 \cdot 10^{-3}$, $\delta A l^3 = 6.0 \cdot 10^{-3}$ and $\delta A l^3 = - 3.0 \cdot 10^{-3}$, respectively. 
The corresponding sets of valid backflow parameter values are then plotted via green dashed, solid and dotted lines, respectively, with the near-intersection denoting the solution. The line equation $\tilde{A}[\tilde{l}]$ for satisfying this condition are found by first-order expansion at the origin via
\begin{equation}
\begin{gathered}
    \delta (A l^3) \approx \delta A \, \tilde{l}^3 + \tilde{A} \, \delta l^3 + \delta A \, \delta l^3 = \\ = (\tilde{A} - 0) 0 + 0 (\tilde{l}^3 - 0) + (\tilde{A} - 0)(\tilde{l}^3 - 0) \rightarrow \\ \rightarrow \tilde{A} = \frac{\delta (A l^3)}{\tilde{l}^3}.
\end{gathered}
\end{equation}
Note that we expand the two-dimensional Taylor series for, both, $\delta A$ and $\delta l^3$ separately, rather than treating the composite term $\delta Al^3$ as a single one-dimensional variable. 
This is so we decouple these variables, in order to find undertake a gradient search in each one. 
Repeating the same procedure, to obtain a non-origin expansion at $\{ A, l \}$ backflow parametrisation, yields
\begin{equation}
    \label{equation:BackflowGradient}
    \tilde{A} = \frac{\delta(A l^3)}{(\tilde{l}^3 - l^3)} + A.
\end{equation}

After plotting all three lines, discussed above, it can be immediately seen that this results in an agreement with the values obtained from the machine learning and direct grid sampling approaches. We believe that the imperfect character of the intersect for the three curves is caused by the small deviations of Eq. \ref{equation:BackflowCondition} observable measurements, but nevertheless, a relatively accurate and useful result is still given. Additionally, the $\delta A l^3$ values not sampled at origin---closer to the final solution---can be seen to yield a closer estimate to the one produced numerically from the machine learning and grid sampling approaches, as expected from a first-order approximation. Hence, it can be suggested to conduct subsequent measurements of Eq. \ref{equation:BackflowCondition} closer to previous intersects.

We also plot the nodal surface sampling distribution corresponding to the optimised parameters $\{ A = -15.83, l = 0.05374 \}$ described in Appendix \ref{appendix:NeuralNetworkDetails} via Eq. \ref{equation:BackflowSignExpression}. This is displayed in Fig. \ref{figure:TestCaseSignSampling} by fixing three of the four particles and moving the remaining one around to calculate the imaginary part $\mathcal{I}(\sigma)$ of the sign. 
The backflow produces sensitive fractal patterns supporting more complex sampling of the nodal surface, which is hypothetically more appropriate for Fermionic systems. The lack of smoothness is probably part of the reason why continuous normalizing flows, and other machine learning techniques, struggle to learn this structure.

\section{Results}

\subsection{Average sign for a finite temperature quantum harmonic oscillator}

Studying the Hamiltonian in Eq.~\ref{equation:QuantumDotHamiltonian} is of general interest, as it models trapped strongly correlated electrons, properties of which have been speculated to yield better understanding of some notable systems, such as room-temperature superconductors\cite{Pickett2023} and materials for controlled nuclear fusion\cite{Dornheim2024}, among others.

\begin{figure}
    \centering
    \includegraphics[width=1.0\columnwidth]{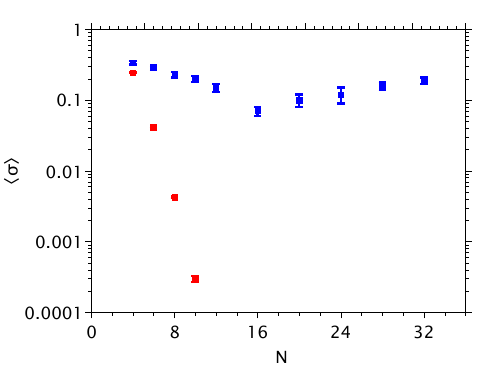}
    \caption{Mean signs, standard deviations from the backflow simulations (blue, dots) and PIMC values\cite{Dornheim2019} of Dornheim (red, squares). Note the dip in our average sign at $N=16$, and the differing gradients above and below this point. Here the two-dimensional trapped electron Hamiltonian in Eq. \ref{equation:QuantumDotHamiltonian} was simulated with modified path integral molecular dynamics codes of Xiong and Xiong\cite{Xiong2022}. Temperature $T = 1.0$ and electron coupling parameter $\gamma = 0.5$ parametrisation was used.}
    \label{figure:ElectrodeCaseDornheimComparisonSign}
\end{figure}

\begin{figure}
    \centering
    \includegraphics[width=1.0\columnwidth]{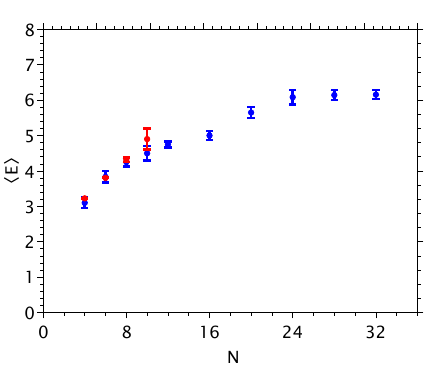}
    \caption{Per-particle mean energies, standard deviations from our backflow simulations (blue, dots) and the PIMC values\cite{Dornheim2019} of Dornheim (red, squares). Note the kink in energy at $N=16$ in our data, and the difference in gradient above and below this point. Here the two-dimensional trapped electron Hamiltonian in Eq. \ref{equation:QuantumDotHamiltonian} was simulated with backflow-modified path integral molecular dynamics codes of Xiong and Xiong\cite{Xiong2022}. Temperature $T = 1.0$ and electron coupling parameter $\gamma = 0.5$ parametrisation was used.}
    \label{figure:ElectrodeCaseDornheimComparisonEnergy}
\end{figure}

\begin{figure}
    \centering
    \includegraphics[width=1.0\columnwidth]{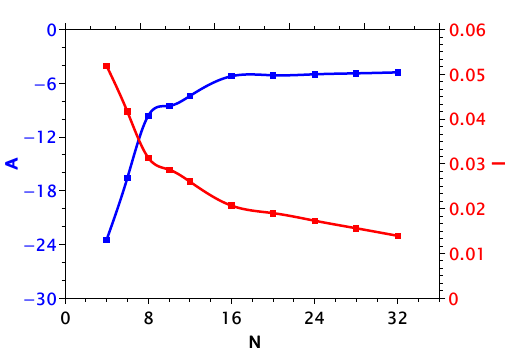}
    \caption{Dependence of the optimised backflow parameters on the particle number. Backflow strength $A$ (blue dots, and interpolated spline) and length $l$ (red dots, and interpolated spline) are shown. Here the two-dimensional trapped electron Hamiltonian in Eq. \ref{equation:QuantumDotHamiltonian} was simulated with backflow-modified path integral molecular dynamics codes of Xiong and Xiong\cite{Xiong2022}. Temperature $T = 1.0$ and electron coupling parameter $\gamma = 0.5$ parametrisation was used.}
    \label{figure:ElectrodeCaseDornheimComparisonBackflow}
\end{figure}

Hence, the backflow sampling procedure, derived in the last section, is utilized to reach particle number regimes too noisy for current approaches. Now, instead of running 128 simulations with different seeds, only a few are run, with the amount of steps conducted corresponding to just enough to reach convergence for each individual case, as in a practical scenario. The resulting sign and per-particle energy values, with their respective standard deviations, are compared, partially, with a previous study\cite{Dornheim2019} and shown in Fig.~\ref{figure:ElectrodeCaseDornheimComparisonSign} and Fig.~\ref{figure:ElectrodeCaseDornheimComparisonEnergy}, respectively, with the backflow parameters displayed in Fig.~\ref{figure:ElectrodeCaseDornheimComparisonBackflow}. Furthermore, it was verified that at the predicted optimal backflow parameters $\delta A l^3 \ll A l^3$ held and it is worth noting that quite often an initial guess around the first curve satisfied this condition already, thus not requiring a second simulation run.

It is immediately visible that the exponential decay of the sign value $\langle \sigma \rangle$ is significantly suppressed, reaching its most severe value of $\sigma = 0.07(1)$ at 16 electrons.

A crossover of behaviour can observed in this region: most clearly in the average sign, but also in the energy and backflow parameters.

This observation is most likely dominated by the shell-closure (in a spin-polarised harmonic trap, the shells are $N = 1, 3, 6, 15, 21$ with $N = 15$ being close to where we observe this crossover).
However this could also be a Wigner localisation effect. We can roughly estimate the respective critical particle number for this using the Lindemann melting criterion, which has been shown\cite{Runge1988} to be relevant for quantum systems as well. It postulates a general condition when quantum or thermal fluctuations overcome the distance between neighbouring particles, hence allowing a phase transition to occur.
It has been recently shown to be valid in two dimensions\cite{Khrapak2020} as well, and states that at the critical point the root-mean-square displacement of the electrons $\langle \xi^2 \rangle$ is related to the characteristic interparticle distance $a$ via a $\sqrt{\langle \xi^2 \rangle} \simeq 0.24 a$ relation. As an estimate, we take $a$ to be the length scale $\sqrt{\frac{\hbar}{m \omega}}$ of the Eq. \ref{equation:QuantumDotHamiltonian} Hamiltonian. 
Since, the quantum dot system, of which the Hamiltonian is defined in Eq. \ref{equation:QuantumDotHamiltonian}, is centred at the origin, the root-mean-square displacement is easily calculated as the average of this value over all of the particles. More specifically,
\begin{equation}
    \langle \xi^2 \rangle = \sum_i \langle |\mathbf{r}_i|^2 \rangle.
\end{equation}

In addition, we also study the sign values and backflow parameters for two additional cases below, even though this energy range is well explored already.

\begin{figure}
    \centering
    \includegraphics[width=1.0\columnwidth]{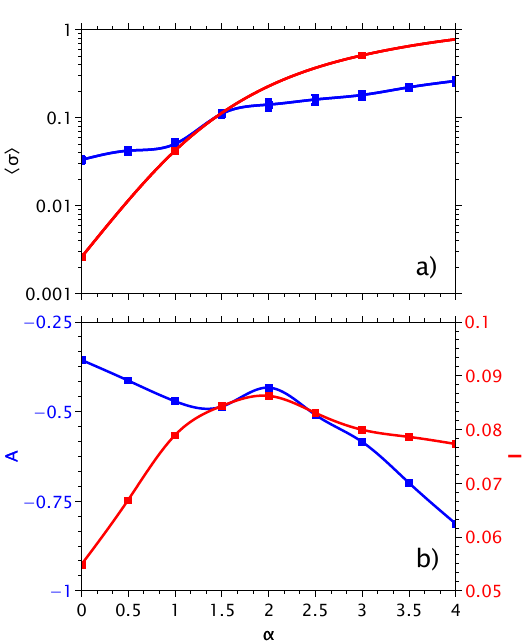}
    \caption{Sign values (a, blue dots and interpolated spline) with the corresponding errors for different interaction powers from the backflow transformed simulation runs. A partial comparison (b, red dots and interpolated spline) with previous study data of Dornheim\cite{Dornheim2019} with no applied backflow is given as well. Interaction power influence (b) on the backflow parameters. The backflow strength $A$ (b, blue dots and interpolated spline) and length $l$ (b, red dots and interpolated spline) are present. Here a modified version of two-dimensional trapped electron Hamiltonian in Eq. \ref{equation:QuantumDotHamiltonian} was simulated with backflow-modified path integral molecular dynamics codes of Xiong and Xiong\cite{Xiong2022}. Temperature $T = 1.0$, electron coupling parameter $\gamma = 0.5$ and $N = 6$ particle number parametrisation was used.}
    \label{figure:ElectrodeCaseInteractionComparison}
\end{figure}

Firstly, to investigate the underlying physics a bit further, the power of the two-body interaction is modified in Eq.~\ref{equation:QuantumDotHamiltonian}, whilst keeping all of the other parameters fixed, via $|\mathbf{r}_i - \mathbf{r}_j| \rightarrow |\mathbf{r}_i - \mathbf{r}_j|^\alpha$, where $\alpha = 1$ corresponds to the Coulomb case mostly studied here. The effect of this is displayed in Fig.~\ref{figure:ElectrodeCaseInteractionComparison} on the sign value and the backflow---$A$ and $l$---parameters. The particle number is set to $N = 6$ as well and, hence, available comparison data\cite{Dornheim2019} is present for the former, with $\alpha = 3$, representative of a dipole-dipole interaction and $\alpha = 0$ to a non-interacting system. It can be immediately seen that the approach developed above is only beneficial for cases with a strong or medium sign problem, basically adding not needed complexity in the weak sign problem case. In all dependencies there is a sort of a change of dependence slope that occurs in the $\alpha \in (1; 2)$ range, possibly indicative again of some sort of a crossover in the physical behaviour. Interestingly, an increase of the interaction order, which is known to decrease the sign problem, actually makes the algorithm in Eq.~\ref{equation:BackflowCondition} predict a stronger backflow. This could be suggestive of the two-dimensional backflow parameter grid, one example of which was displayed in Fig.~\ref{figure:TestCaseOptimalBackflowParameters}, being too flat, most realistically caused from the gradient being negligible due a significant increase of the $\sigma$ sign value not being present from variations in $A$ and $l$ values.

\begin{figure}
    \centering
    \includegraphics[width=1.0\columnwidth]{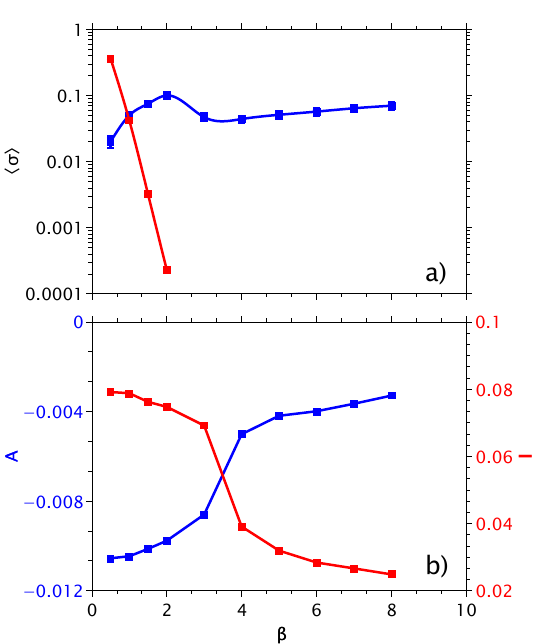}
    \caption{Sign values (a, blue dots and interpolated spline) with the corresponding errors for different temperatures from the backflow transformed simulation runs. A partial comparison (a, red dots and interpolated spline) with previous study data of Dornheim\cite{Dornheim2019} with no applied backflow is given as well. Temperature influence (b) on the backflow parameters. The backflow strength $A$ (b, blue dots and interpolated spline) and length $l$ (b, red dots and interpolated spline) are present. Here the Hamiltonian in Eq. \ref{equation:QuantumDotHamiltonian} was simulated with backflow-modified path integral molecular dynamics codes of Xiong and Xiong\cite{Xiong2022}. Electron coupling parameter $\gamma = 0.5$ and $N = 4$ particle number parametrisation was used.}
    \label{figure:ElectrodeCaseTemperatureComparisonBackflow}
\end{figure}

Secondly, we also vary the inverse temperature $\beta = T^{-1}$ with the parameters above and find a significantly reduced sign problem too. This is displayed in Fig.~\ref{figure:ElectrodeCaseTemperatureComparisonBackflow}. Surprisingly, the magnitude of this effect has quite complex behaviour. 
The worst performance is present at the high temperature regime---again we think we are introducing noise into a situation with no sign problem. 
There is some sort of crossover in the physical behaviour at an intermediate temperature case where the mitigation performance of the backflow takes a small dip before recovering as the system becomes colder.

Generally, in all three sign problem cases of particle number, temperature and interaction type we find that, roughly, the length scale $l$ is the same---ranging from $l \sim 0.01$ with severe sign problem, to $l \sim 0.1$ where there is no such issue. 
It is the strength $A$ of the backflow that varies by orders of magnitude depending on the parameter range, $A \sim -10$ for particle number, $A \sim -0.1$ for interaction type and $A \sim -0.001$ for temperature variations. 
Additionally, each of the three cases see an approximate fourfold decrease of the strength $A$ as the sign problem becomes worse.
The significant magnitude changes of this parameter can be possibly explained by the major differences in bead cluster---converged PIMD sampling configuration---sizes for each of the three variables. Specifically, this has been observed before\cite{Dornheim2019} when visualising PIMC simulation data.

\section{Conclusion}

We have shown that the simple hydrodynamic backflow transformation significantly reduces the Fermion sign problem in path integral molecular dynamic simulations of a harmonically trapped two-dimensional electron gas at finite temperature. These simple wavefunctions\cite{Kruger2008, Kaplis2016} can support a rich, fractal, nodal structure, and by scaling arguments have been suggested to be good models for the Fermi liquid.
We found that directly incorporating the quasi-position coordinate transform of the backflow wavefunction into a path integral molecular dynamics code led to improved convergence and unbiased estimators of the energy. We emphasize that this can be implemented on any other software of this type with minor effort.

Our first attempt to optimise parameters of these backflows was a continuous normalizing flow approach which was found to reduce the relative error of the total energy, approximately, three times at medium sign severity, from $\langle \sigma \rangle = 0.2(1)$ to $\langle \sigma \rangle = 0.5(1)$ respectively. Overall, whilst we could learn directly from the path integral molecular-dynamic samples, the method was inefficient, numerically unstable and brittle.

We therefore developed a semi-analytic analogue to directly optimise the backflow parameters with the aid of a closed-form solution for first order improvement of the sign, and used this as the gradient for a standard optimisation algorithm. The resulting transformation agreed with the previous continuous normalizing flow approach. The computationally limiting factor of our semi-analytic method was the $O(N^3)$ calculation of the Jacobian, stemming from the coordinate transformation of the backflow. Approximate Jacobians, using more compute time, or even just a careful and optimised algorithm, may improve this limitation and enable scaling beyond 32 particles.

Overall we reduced the sign-problem by orders of magnitude in the more severe cases ($\langle \sigma \rangle = 0.07(1)$ at minimum). The total energy of our backflow-transformed system matched with available data. We successfully calculated the energy of a 32 electron system, where previous calculations at this couplings only stretched to 10 electrons. A crossover in the physics at 16 trapped particles was identified directly from the energy versus particle number curve, and a similar change in the average sign, among other parameter variations.
The Lindemann criteria for Wigner crystallisation was calculated to be at $N = 18$ electrons; but also the two-dimension electron shell is filled at $N = 15$ electrons.
Overall the optimal decay length for electron backflow was found to be $l \sim 0.01 - 0.1$, while the backflow strength varied 4 orders of magnitude with $A \sim 10^{-3} - 10^{1}$ values.

As a final note, though the Hamiltonian in Eq. \ref{equation:QuantumDotHamiltonian} is abstract, it is often studied as a model\cite{Reimann2002} for two dimensional quantum dots, which have a number\cite{Kumar2020} of technical applications, such as in\cite{Zhang2018} supercapacitors. During the writing of this article we found that the parameters used here describe\cite{Neto2007, Tiras2013, Gttinger2010} graphene quantum dots quite well, but this has been left to be studied in future work.

\section{Acknowledgement}

The authors gratefully acknowledge fruitful discussions with Matthew Foulkes, Yunuo Xiong and Frank Kruger.
J.M.F. is a Royal Society University Research Fellow (URF-R1-191292). I.V. is supported by a Royal Society doctoral studentship from the same grant. The authors gratefully acknowledge access to computing resources provided by the Imperial College Research Computing Service\cite{ImperialHPC}. Derivations in this work were carried out with the help of the Mathematica\cite{Mathematica} symbolic manipulation software.

\appendix

\section{Neural Network Approach}

\label{appendix:NeuralNetworkDetails}

\begin{figure}
    \centering
    \includegraphics[width=1.0\columnwidth]{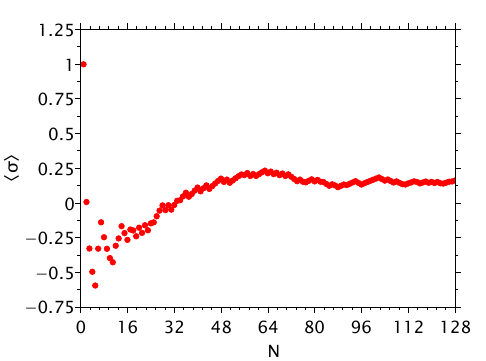}
    \caption{Convergence of the average sign $\langle \sigma \rangle$ with respect to the size $N$ of the backflow untransformed simulation dataset. Here the Hamiltonian in Eq. \ref{equation:QuantumDotHamiltonian} was simulated with path integral molecular dynamics codes of Xiong and Xiong\cite{Xiong2022} for generating learning data. Temperature $T = 1.0$, electron coupling parameter $\gamma = 0.5$ and $N = 4$ particle parametrisation was used.}
    \label{figure:TestCaseSignConvergence}
\end{figure}

Each of the PIMD simulations, used for learning, were run separately with a unique random number generator seed and a single sign value and particle bead configuration extracted for the learning process. This was done in order to remove auto-correlation effects and random generator seed dependence as much as possible and yielded an average sign value of $\langle \sigma \rangle = 0.2(1)$ for the backflow untransformed case, as seen in Fig.~\ref{figure:TestCaseSignConvergence}. Additionally, an energy value of $E = 13(1)$ was obtained, hence both observable values being in agreement with the original study\cite{Xiong2022} as well. Note that the optimal size of the simulation dataset was estimated to be $N = 128$ by looking at the convergence of the average sign $\langle \sigma \rangle$, displayed in Fig.~\ref{figure:TestCaseSignConvergence}, with respect to its size.

In order to further perform minimization for the learning process, a step size of $10^{-3}$ was used together with the \textit{AdamW} solver and it was stopped when the relative difference of the cost function reached $10^{-5}$ between steps. The learning process itself required a careful approach, since any significant instability would just produce a backflow representing a constant coordinate shift of the whole simulation independent of the sign, meaning that no nodal structure had been extracted. Hence, in order to remedy this, the neural networks were first trained on a very small dataset size of $N_s = 4$ and, further, the weights of these networks were used as an initial guess to converge them for higher dataset sizes of, first, $N_s = 16$ and, finally, $N_s = 128$ sizes. Additionally, the neural networks $\{ A_n(\sigma, T) \}$ and $\{ l_n(T) \}$ themselves were found to be of satisfactory composition with one hidden linear class layer and four neurons, enough to describe a few cycles of a \textit{sin(x)} trigonometric function with relative deviations within a percent. More complex forms were found to introduce instabilities in the learning process and not yield a better description of the backflows at the same time.

\begin{figure}
    \centering
    \includegraphics[width=1.0\columnwidth]{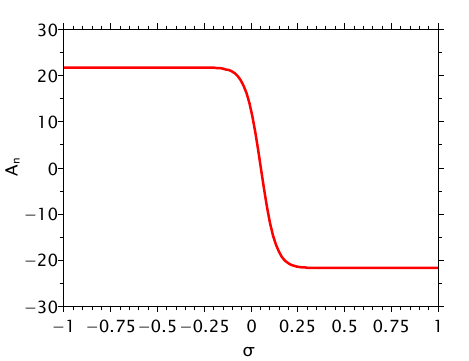}
    \caption{Learned backflow transformation strength scale $A$ dependence on the sign $\sigma$ in the case of a single backflow. Here the Hamiltonian in Eq. \ref{equation:QuantumDotHamiltonian} was simulated with path integral molecular dynamics codes of Xiong and Xiong\cite{Xiong2022} for generating learning data. Temperature $T = 1.0$, electron coupling parameter $\gamma = 0.5$ and $N = 4$ particle parametrisation was used.}
    \label{figure:TestCaseBackflowStrength}
\end{figure}

In the case of a single backflow, the learned strength scale is displayed in Fig.~\ref{figure:TestCaseBackflowStrength} with respect to the sign. It can be seen that the zero intercept of the strength scale approximately crystallizes out the average sign value of the untransformed simulation $\langle \sigma_0 \rangle = 0.2(1)$, as previously discussed and expected. The strengths can be seen to be dependent on the sign with a form similar to the \textit{tanh} function, which is most likely due to the fact that the sign values of the individual samples tend to be either $\sigma \approx \pm 1$ and not often anything in between.

Moving on, this backflow was implemented on top of the stochastic path integral ring polymer molecular dynamics software. Specifically, the \textit{RelativeDistance} and \textit{Distance} functions were modified using Eq.~\ref{equation:BackflowTransformation} in order to make the simulation sample the backflow transformed coordinate space with a strength of $A_1(T = 1.0, \sigma = 1.0) = -15.83$ and scale $l_1(T = 1.0) = 0.05374$. This did require a decrease of the time step to $h = 0.00125$---half its original value, otherwise most of the simulation runs would fail. This yielded an average sign value of $\langle \sigma \rangle = 0.5(1)$, in a very similar fashion as in Fig.~\ref{figure:TestCaseSignConvergence}, and an energy value of $E = 13(1)$, hence agreeing with the energy value of the untransformed simulation dataset and showing an increased average sign value, signifying a better sampling of the Fermionic nodal surface. Note that, since the neural network $\{ A_n(\sigma, T) \}$ has no bounds on $\sigma$, it was never expected for the simulation to yield the perfect average sign value of $\langle \sigma \rangle$ value of unity, but rather depict how to achieve its maximum value instead, as discussed previously.

The above learning process was repeated for the case of four backflows. One of these resulted in having negligible impact and the rest were of practically identical length $l_{1,2,3}(T = 1.0) \approx 0.0346$ and strength $A_{1,2,3}(T = 1.0, \sigma = 1.0) \approx -13.8$ scale characterization, hence showing that, in this case, a single backflow function is enough for obtaining the best sampling description of the Fermionic nodal surface. This fact is further supported by comparing both simulation strengths via $A_1 l_1^3 \approx A_1 l_1^3 + A_2 l_2^3 + A_3 l_3^3 + A_4 l_4^3$ which shows that both results yield the same coordinate transformation effect. This agrees with previous VMC studies\cite{Ros2006} that used an arbitrary $8^\text{th}$ order polynomial and, by energy minimization, found a result very similar to the form used here. In an interesting manner, the values of these four backflows yielded a slightly lower average sign value of $\langle \sigma \rangle = 0.4(1)$, most probably caused by the noise from over-representation, in comparison with the single backflow case. A further direct grid sampling of varying length and strength scales is displayed in Fig.~\ref{figure:TestCaseBackflowStrengthAndScale}. It can be reaffirmed that the above stated values, represented by a blue dot, do indeed represent a maximum on the $\langle \sigma \rangle (A_1, l_1)$ map and it is interesting to note that at certain values the average sign value can actually become small and, therefore, be more costly to evaluate. It is expected that the small deviation from the simulation values, present in the figure, is caused by finite size effects, since $64$ square bins were used here with approximately $128$ simulation values of the average sign $\langle \sigma \rangle$ for each one. This was of identical amount to the previous discussion, summarised by Fig.~\ref{figure:TestCaseSignConvergence}, in order to achieve convergence.

\section{Backflow Energy in Momentum Space}

\label{appendix:BackflowMomentumSpace}

Here we show for reference, in one spatial dimension, how the exchange energy term of the hydrodynamical backflow in Eq. \ref{equation:FirstOrderEnergy} is converted into coordinate space. This is done via the Fourier transform in two dimensions just like in Eq. \ref{equation:PlaneWaveTransform} but now two momenta components $k_n = \frac{\pi \, n}{L}$ and $k_m = \frac{\pi \, m}{L}$ are considered corresponding to each particle. This results in a $x_1$ and $x_2$ dependency via
\begin{equation}
\begin{gathered}
    \sum_{n=1}^{\infty} \sum_{m=1}^{\infty} [k_m^3 - k_n k_m^2 - k_n^2 k_m + k_n^3] \, e^{2 i x_1 k_m} e^{2 i x_2 k_n} = \\ = -\frac{\pi ^3 \csc ^4\left(\frac{\pi  x_1}{L}\right)}{8 L^3} - \frac{\pi ^3 \csc ^4\left(\frac{\pi  x_2}{L}\right)}{8 L^3} - \\ - \frac{\pi ^3 \cos \left(\frac{2 \pi  x_1}{L}\right) \csc ^4\left(\frac{\pi  x_1}{L}\right)}{16 L^3} - \\ - \frac{\pi ^3 \cos \left(\frac{2 \pi  x_2}{L}\right) \csc ^4\left(\frac{\pi  x_2}{L}\right)}{16 L^3} + \\ + \frac{i \pi ^3 \cot \left(\frac{\pi  x_2}{L}\right) \csc ^4\left(\frac{\pi  x_1}{L}\right)}{8 L^3} + \\ + \frac{i \pi ^3 \cot \left(\frac{\pi  x_1}{L}\right) \csc ^4\left(\frac{\pi  x_2}{L}\right)}{8 L^3} - \\ - \frac{i \pi ^3 \sin \left(\frac{\pi  \left(x_1+x_2\right)}{L}\right) \csc ^3\left(\frac{\pi  x_2}{L}\right) \csc ^3\left(\frac{\pi  x_1}{L}\right)}{16 L^3} + \\ + \frac{i \pi ^3 \cos \left(\frac{2 \pi  x_1}{L}\right) \cot \left(\frac{\pi  x_2}{L}\right) \csc ^4\left(\frac{\pi  x_1}{L}\right)}{16 L^3} + \\ + \frac{i \pi ^3 \cos \left(\frac{2 \pi  x_2}{L}\right) \cot \left(\frac{\pi  x_1}{L}\right) \csc ^4\left(\frac{\pi  x_2}{L}\right)}{16 L^3}.
\end{gathered}
\end{equation}
Finally, we remove the periodic constraint which is not present, for example, in the quantum dot system above. This is done by applying $L \gg x_1, x_2$ and results in
\begin{equation}
    \frac{i L^2 \left(x_1+x_2\right) \left(3 x_1^2-4 x_2 x_1+3 x_2^2\right)}{16 \pi ^2 x_1^4 x_2^4}.
\end{equation}
Note that it can be easily seen from Eq. \ref{equation:FirstOrderEnergy} that the box length $L$ drops out of the final energy and, hence, sign expression in Eq. \ref{equation:BackflowSignExpression}.

\bibliography{sources}

\end{document}